\documentstyle[12pt]{article}
\author{Walter Simon\thanks{Present address: Institut f\"ur theoretische
Physik der Univ. Wien, Boltzmanngasse 5, A-1090 Wien, Austria.
E-mail: simon@pap.univie.ac.at}\\
Dept. of Mathematics\\ The University of New England\\
Armidale NSW 2351, Australia\\and\\
Centre for Mathematics and its Applications\\
Australian National University\\
Canberra ACT 0200, Australia}
\title{Nuts have no hair}
\date{}
\begin{document}
\maketitle
\begin{abstract}
We show that the Riemannian Kerr solutions
are the only Riemannian, Ricci-flat and asymptotically flat
${\rm C}^{2}$-metrics $g_{\mu\nu}$
on a 4-dimensional complete manifold ${\cal M}$ of topology
${\rm R}^{2} \times {\rm S}^{2}$
which have (at least) a 1-parameter group of periodic isometries
with only isolated fixed points ("nuts") and with orbits of bounded
length at infinity.
\end{abstract}
\newpage
The relevance of instantons, (here understood as being regular, real,
Riemannian, Ricci-flat manifolds), in quantum gravity \cite{SH}-\cite{MP}
has stimulated interest in theorems on the (non-) existence
in particular of periodic such solutions and of solutions with isometries.
It has been shown under rather general conditions that there are no non-trivial
instantons on ${\rm R}^{4}$ and on ${\rm R}^{3} \times {\rm S}$
\cite{WMS}. Known examples include the ${\rm R}^{2} \times {\rm S}^{2}$
Kerr-NUT instanton \cite{GH1,MP,GH2}
which (like many others) has been found by
``Euclideanizing'' the corresponding Lorentzian solution \cite{RK},
and the adaption of Lorentzian uniqueness (``no-hair'') theorems
has been discussed as well \cite{AL}. The different character of the
Riemannian case (the absence of horizons and ergospheres and
the existence of singularity-free solutions) require and suggest,
however, alternative approaches to the uniqueness problem.
Based on a characterization of the Lorentzian Kerr metric in terms of
complex quantities \cite{WS} which become real in the Riemannian case
and also satisfy generalizations of ``Israel''-type identities \cite{IR}
we have obtained the following result. (We abbreviate ``Riemannian'' by
``Riem.'' and ``Lorentzian'' by ``Lor.'' henceforth. Greek indices go
from 0 to 3).\\ \\
{\it Theorem.}
The Riem. Kerr solutions
are the only Riem., Ricci-flat and asymptotically flat
${\rm C}^{2}$-metrics $g_{\mu\nu}$
on a 4-dimensional complete manifold ${\cal M}$ of topology
${\rm R}^{2} \times {\rm S}^{2}$
which have (at least) a 1-parameter group of periodic isometries
with only isolated fixed points and with orbits of bounded length at infinity.
\\

Introductory material and 2 Lemmas will precede the proof. Details of parts
of our analysis and extensions thereof will be given elsewhere.

The condition of Ricci flatness ($R_{\mu\nu} = 0$) implies that
$g_{\mu\nu}$ and the Killing field $\xi^{\mu}$ corresponding to the isometry
$\mu_{\tau}$ ($\tau$ is the group parameter) are analytic in harmonic
coordinates \cite{HM}. The set ${\cal L}$ of fixed points of $\mu_{\tau}$ has
the following structure \cite{GH1}. At every $q \in {\cal L}$  the differential
$\mu_{\tau \ast}$ leaves invariant two 2-dimensional orthogonal subspaces
$T_{q}^{+}$ and $T_{q}^{-}$ of the tangent space $T_{q}$. If $\mu_{\tau \ast}$
acts as the identity on one of $T_{q}^{+}$ or $T_{q}^{-}$ there is a 2-surface
of fixed points called "bolt" which we exclude by assumption. If $q$ is
isolated it is called a "nut" after the Taub-NUT metric \cite{TN}.
In this case $\mu_{\tau \ast}$ acts as rotations
in each of $T_{q}^{+}$ and $T_{q}^{-}$ with periods $\tau^{\pm} =
 2\pi/\kappa^{\pm}$ (the smallest values of $\tau$ such that
$\mu_{\tau \ast} X^{\pm} = X^{\pm}$ for $X^{\pm} \in T_{q}^{\pm}$).
$\kappa^{+}$ and $\kappa^{-}$ are also the skew eigenvalues of
$\nabla_{\mu}\xi_{\nu}$ in an orthonormal frame and called "gravities" of the
nut. As $\mu_{\tau}$ is assumed to
be periodic, there is a (smallest) $\tau^{0}$ such that
$\mu_{\tau^{0} \ast} X = X$ for all $X \in T_{q}$ which implies that
$\tau^{+} p^{+} = \tau^{0} = \tau^{-} p^{-}$ for relative
prime integers $p^{+}$ and $p^{-}$. Since $\mu_{\tau}$ commutes with the
exponential map, i.e. $\exp(\mu_{\tau \ast} X) = \mu_{\tau}(\exp X)$, the
period of $\mu_{\tau \ast} X $ at $q$ equals the periods of the orbits through
all points of a geodesic emanating from $q$ with tangent vector $X$ (at least)
as long as the exponential map is non-singular.
\vfill\eject
Let $\lambda = \xi^{\mu}\xi_{\mu}$ denote the norm and
$\omega_{\mu} = \epsilon_{\mu\nu\sigma\tau}\xi^{\nu}\nabla^{\sigma}\xi^{\tau}$
the twist of $\xi^{\mu}$. ($\epsilon_{\mu\nu\sigma\tau}$ is antisymmetric and
$\epsilon_{0123} = (det~ g)^{1/2}$).
By Ricci flatness, $\nabla_{\mu}(\lambda^{-2}\omega^{\mu}) = 0$. Hence the
"nut charge" \cite{GH1}
\begin{equation}
\label{n}
m_{i}^{\ast} = \frac{1}{8\pi} \int_{{\cal S}_{i}}\lambda^{-2}
\omega_{\mu}dS^{\mu} = \frac{\pi}{2\kappa^{+}_{i}\kappa^{-}_{i}}
\end{equation}
is independent of the compact 3-surface ${\cal S}_{i}$ which
encloses the nut $n_{i}$ and does not intersect others. The surface element
$dS^{\mu}$ points outwards. The second part of (\ref{n}) follows by
Taylor-expanding $\xi^{\mu}$ at $n_{i}$ and by shrinking ${\cal S}_{i}$ to
$n_{i}$.

In Lemma 1 and in the Theorem we adopt a standard definition
of asymptotic flatness (AF) \cite{MP,GPR} and require ${\cal M}$ minus a
compact set to be diffeomorphic to ${\rm R}^{+} \times {\rm S}
 \times {\rm S}^{2}$ and the
metric and its first and second derivatives to go to the flat metric and its
derivatives, with the usual $1/r-$ falloff in coordinates adapted to the
isometry, i.e. $\partial_{\sigma}g_{\mu\nu} = 0$. The definition implies that
at infinity all orbits $\mu_{\tau}$ have the same length  which we call
$l_{\infty}$. (We remark, however, that the limit of the length function may
be discontinuous when the limiting orbit is approached via orbits which wind
repeatedly around the large ${\rm S}^{2} \times {\rm S}$-surfaces of constant
distance from a nut). In Lemma 2 we will require "local asymptotic
flatness" (ALF) with the cyclic group Z \cite{MP,GPR}. In this setting we
can define the "dual mass" $m^{\ast}$ \cite{RS} by considering the integral
in (\ref{n}) over the asymptotic region. We remark that
$({\cal M},g_{\mu\nu})$ is AF iff it is ALF and $m^{\ast} = 0$. $\xi^{\mu}$ is
normalized such that $\lambda \rightarrow 1$ at infinity.\\ \\
{\it Lemma 1.} Under the requirements of the Theorem ${\cal M}$ has precisely 2
nuts $n_{1}$ and $n_{2}$ whose "gravities" $\kappa_{1}^{\pm}$ and
$\kappa_{2}^{\pm}$ satisfy $\kappa_{1}^{+} = \kappa_{2}^{+}$ and
$\kappa_{1}^{-} =-\kappa_{2}^{-}$. (The choice of the labels $+$ and $-$ is a
convention). Moreover, we have
$l_{\infty} \ge \min(2\pi/\kappa^{+}, 2\pi/\kappa^{-})$
where $\kappa^{\pm} = |\kappa_{1}^{\pm}| = |\kappa_{2}^{\pm}|$. \\ \\
{\it Proof.} As ${\cal M}$ has topology ${\rm R}^{2} \times {\rm S}^{2}$, it
has Euler number $\chi = 2$ and signature $\tau = 0$. In the absence of bolts
and using AF, the index theorem implies that $\chi$ is equal to the number of
nuts and $3\tau = \kappa_{1}^{+}/\kappa_{1}^{-} + \kappa_{1}^{-}/\kappa_{1}^{+}
+ \kappa_{2}^{+}/\kappa_{2}^{-} +\kappa_{2}^{-}/\kappa_{2}^{+}$. (See
\cite{BAB} for the compact case and \cite{GPR,APS} regarding boundary terms).
Together with (\ref{n}) and $m_{1}^{\ast} + m_{2}^{\ast} = m^{\ast} = 0$
we obtain the first part of the lemma.

As ${\cal M}$ is not compact there is (at least one) $X_{1} \in T_{n_{1}}$
and (at least one) $X_{2} \in T_{n_{2}}$ such that $\gamma_{1} =
\exp(tX_{1})$ and $\gamma_{2} = \exp(tX_{2})$, $t \in (0,\infty)$
approach infinity as minimizing (radial) geodesics \cite{KN}.
As families $\mu_{\tau}(\gamma_{1})$ and
$\mu_{\tau}(\gamma_{2})$ of such geodesics diverge in the asymptotic region,
the exponential map remains non-singular in the limit. Hence $l_{\infty}$
equals the periods of $\mu_{\tau \ast}X_{1}$ and $\mu_{\tau \ast}X_{2}$ which
can be $\tau^{\pm}$ or $\tau^{0} \ge \tau_{\pm}$. Thus the lemma holds.
\vfill\eject
{}From $R_{\mu\nu} = 0$, $\omega_{\mu}$ is curl-free, i.e.
$\nabla_{[\mu}\omega_{\nu]} = 0$. As ${\cal M}$ is simply connected,
$\nabla_{\mu}\omega = \omega_{\mu}$ defines a scalar
field $\omega$ globally and up to a constant which we choose such that
$\omega$ vanishes at infinity. We also define
${\cal E}_{\pm} = \lambda \pm \omega$,
$\mu = \frac{1}{2}(\nabla_{\mu}\xi_{\nu})(\nabla^{\mu}\xi^{\nu})$ and
$\nu = \frac{1}{4}\epsilon_{\mu\nu\sigma\tau}(\nabla^{\mu}\xi^{\nu})
(\nabla^{\sigma}\xi^{\tau})$ which satisfy, again from $R_{\mu\nu} = 0$,
\begin{equation}
\label{dde}
\Box {\cal E}_{\pm} = 4(\mu \pm \nu) = \lambda^{-1} \nabla_{\mu}{\cal E}_{\pm}
\nabla^{\mu}{\cal E}_{\pm} \ge 0.
\end{equation}
The maximum principle and the asymptotic conditions imply
${\cal E}_{\pm} < 1$ and hence
${\cal E}_{\pm} = - {\cal E}_{\mp} + 2\lambda \ge - {\cal E}_{\mp} > -1$.

To simplify what follows we now foliate ${\cal M} \setminus {\cal L}$ by the
orbits of $\mu_{\tau}$ \cite{GH1,RG}. We obtain a manifold
$({\cal N},\gamma_{ij})$ where $\gamma_{ij}$ is the pullback of
$\gamma_{\mu\nu} = \lambda g_{\mu\nu} - \xi_{\mu}\xi_{\nu}$.
(Tensors on ${\cal N}$ carry latin indices).
We denote by $D_{i}$ and $R_{ij}$ the covariant derivative and the Ricci
tensor with respect to $\gamma_{ij}$ and introduce
$w_{\pm}=(1 + {\cal E}_{\pm})^{-1}(1 - {\cal E}_{\pm}),~
\Theta = 1 - w_{+}w_{-},~A_{i} = \frac{1}{2}(w_{+}D_{i}w_{-}-w_{-}D_{i}w_{+})$
and ${\cal D}_{\pm}^{i} = \Theta^{-1}D^{i} \mp 2\Theta^{-2}A^{i}$.
Since $|{\cal E}_{\pm}| < 1$ we have $0 < w_{\pm} < \infty$ and $\Theta > 0$.
On ${\cal N}$ the condition $R_{\mu\nu} = 0$ reads
\begin{equation}
\label{ddw}
D_{i}{\cal D}^{i}_{\pm} w_{\pm} = 0
\end{equation}
\begin{equation}
\label{ricc}
R_{ij} = 2\Theta^{-2}D_{(i}w_{-}D_{j)}w_{+}.
\end{equation}
When $({\cal M}, g_{\mu\nu})$ is ALF, $({\cal N}, w_{\pm}, \gamma_{ij})$ is
asymptotically flat in a standard sense (compare \cite{BS}).

In coordinates $r = \Re - m$ where $\Re$ is the radial "Boyer-Lindquist"-
coordinate (equ. (2.13) of \cite{BL}) the Riem. Kerr-NUT metric reads
\begin{equation}
\label{wkerr}
w_{\pm} = m_{\pm}(r \pm \alpha~ cos\theta)^{-1},
\end{equation}
\begin{eqnarray}
\label{gkerr}
\lefteqn{\gamma_{ij}dx^{i}dx^{j} =  (r^{2}-m_{+}m_{-}-\alpha^{2})^{-1}
(r^{2}-m_{+}m_{-}-\alpha^{2}cos^{2}\theta)dr^{2} +} \nonumber\\ & & +
 (r^{2}-m_{+}m_{-}-\alpha^{2}cos^{2}\theta)d\theta^{2}+
 (r^{2}-m_{+}m_{-}-\alpha^{2})sin^{2}\theta d\phi^{2}.
\end{eqnarray}
Here $m = \frac{1}{2}(m_{+} + m_{-})$ and
$m^{\ast} = \frac{1}{2}(m_{+} - m_{-})$ are the mass and
the dual mass and $\alpha$ is another real constant. For $m^{\ast} = 0$ this
is the Riem. Kerr metric for which $\xi^{\mu} = \partial/\partial \tau$ has 2
nuts at $r = \sqrt{m^{2} + \alpha^{2}},~\theta = 0~{\rm and}~\theta = \pi$.
In the Riem. Schwarzschild case ($m^{\ast} = \alpha = 0$) this vector has a
bolt at $r = m$. For the Riem. Kerr metric Kruskal-like coordinates can be
obtained by "Euclideanizing" (3.8) of \cite{BL}.

Our characterization  involves the pairs of quantities
\begin{equation} \label{k}
k_{\pm}^{4} = D^{i}w_{\pm} D_{i}w_{\pm},
\end{equation}
\begin{equation} \label{b}
 B^{\pm}_{ij} =
4\Theta^{-2}{\cal C}[D_{i}D_{j}w_{\pm} - (3w^{-1}_{\pm} + \Theta^{-1} w_{\mp})
D_{i}w_{\pm} D_{j}w_{\pm}],
\end{equation}
where ${\cal C}$ denotes the trace-free part and
\vfill\eject
\begin{equation} \label{c}
C^{\pm}_{ijk} = 4\Theta^{-2}(D_{i}D_{[j}w_{\pm}D_{k]}w_{\pm} -
  \gamma_{i[j} u^{\pm}_{k]}),
\end{equation}
where
\begin{equation} \label{u}
u^{\pm}_{k} = \gamma^{ij} D_{i} D_{[j} w_{\pm} D_{k]} w_{\pm}.
\end{equation}
On sets where $k_{\pm}^{4} \ne 0$ (\ref{ddw}) and (\ref{ricc}) imply,
for each $\alpha \in {\rm R}$,
\begin{eqnarray}
\label{ddkw}
D_{i}{\cal D}_{\pm}^{i}\frac{k^{\alpha + 1}_{\pm}}{w^{\alpha}_{\pm}}
& = &\alpha (\alpha + 1) \frac{k^{\alpha - 1}_{\pm}}
{\Theta w^{\alpha}_{\pm}}
(D_{i} k_{\pm} - \frac{k_{\pm}}{w_{\pm}}D_{i}w_{\pm})
(D^{i} k_{\pm} - \frac{k_{\pm}}{w_{\pm}}D^{i}w_{\pm}) +
\nonumber \\
& & +\frac{\alpha + 1}{16} \frac{k^{\alpha - 7}_{\pm}}{w^{\alpha}_{\pm}}
\Theta^{3} C_{ijk}^{\pm} C_{\pm}^{ijk}.
\end{eqnarray}
For $\alpha \ge 0$, the r.h. sides of (\ref{ddkw}) are non-negative and
for $\alpha = 3$ they can be written as
$\frac{1}{8} \Theta^{3} w^{-3}_{\pm}B_{ij}^{\pm}B^{ij}_{\pm}$.

When $\xi^{\mu}$ is hypersurface-orthogonal ($\omega = 0$) the objects
$B^{\pm}_{ij}$ coincide and are, by virtue of (\ref{ddw}) and (\ref{ricc}),
equal to certain
functions $f_{\pm}(\lambda)$ times the Ricci tensors $R_{ij}^{\pm}$ with
respect to the metrics $\gamma_{ij}^{\pm}  =
\frac{1}{16} \lambda^{-1}(1 \pm \lambda^{1/2})^{4}\gamma_{ij}$.
Likewise, for $\omega = 0$ each of $C_{ijk}^{\pm}$ reduces to the Cotton tensor
which characterizes conformal flatness.
The corresponding characterizations of the Lor. Schwarzschild metric
and the restriction of (\ref{ddkw}) for certain values of $\alpha$ were
employed in uniqueness proofs \cite{IR, MRS}. In the general case
$B_{ij}^{\pm}$, $k_{\pm}$ and $C_{ijk}^{\pm}$  have complex Lor.
counterparts $B_{ij}$, $k$ and $C_{ijk}$ which have analogous properties.
 The latter two quantities have been employed
in local characterizations of the Kerr metric among the AF
 ones \cite{WS} and of a larger class of metrics if the asymptotic
assumption is dropped \cite {ZP}. The methods of these papers can be
straightforwardly applied in the Riem. case and yield the following
result. \\ \\
{\it Lemma 2.} An ALF ${\rm C}^{2}$-solution $(w_{\pm}, \gamma_{ij})$
of (\ref{ddw}) and (\ref{ricc}) is isometric to a Riem. Kerr-NUT metric iff it
satisfies one of (\ref{kw}), (\ref{b0})
or (\ref{c0}) (a pair of equations in each case) in a neighbourhood ${\cal U}$
of a point of ${\cal N}$:
\begin{equation} \label{kw}
\mbox{Either}~~k_{+} = \sigma_{+} w_{+}~~\mbox{or}~~ w_{+} = 0, ~
\mbox{and either}~~k_{-} = \sigma_{-} w_{-} ~~ \mbox{or} ~~ w_{-} = 0,
\end{equation}
where $\sigma_{\pm} > 0$  are constants.
\begin{equation} \label{b0}
B^{\pm}_{ij}=0.
\end{equation}
\begin{equation} \label{c0}
C^{\pm}_{ijk}=0.
\end{equation}\\
{\it Proof.} Degenerate cases in which either $w_{+}$
and $w_{-}$ are functionally related or one of $w_{\pm}$ vanishes on ${\cal U}$
are easily disposed of. In the generic case, from (\ref{ddw}) and
(\ref{ddkw}), (\ref{kw}) implies (\ref{b0}) and (\ref{c0}).
Conversely, (\ref{kw}) follows either by inserting (\ref{b}) into
$B_{ij}^{\pm}w_{\pm}^{j} =0$, using also (\ref{ddw}), or
from $C_{ijk}^{\pm} = 0$ and the ALF conditions as in the Lor. case
\cite {WS,ZP}.
\vfill\eject
The Riem. Kerr-NUT metric in the form (\ref{wkerr}), (\ref{gkerr}) is easily
seen to satisfy (\ref{kw}). To show the converse we essentially follow
\cite {ZP} and define the vector field
\begin{equation} \label{l}
l_{i}= -\frac{1}{2}\sigma^{-4}_{-} \sigma^{-4}_{+} w^{-3}_{-}w^{-3}_{+}~
\Theta~ \epsilon_{ijk} (D^{j}w_{-})(D^{k}w_{+})
\end{equation}
which, from (\ref{ddw}) and (\ref{b0}), is hypersurface-orthogonal
($\epsilon_{ijk}l^{i}D^{j}l^{k} =0$) and Killing ($D_{(i}l_{j)}=0).$
Hence there exists a function $r_{0}$ on ${\cal U}$ such that $l^{i} =
\partial/\partial r_{0}$ and the metric coefficients in the coordinates
$r_{0}$ and $r_{\pm}= \frac{1}{2}
(\sigma^{-2}_{+}w^{-1}_{+} \pm \sigma^{-2}_{-}w^{-1}_{-})$ are independent
of $r_{0}$. Moreover, from (\ref{kw}) and (\ref{l}) the metric is diagonal in
 these coordinates and
\begin{equation} \label{gr}
\gamma^{r_{+}r_{+}} + \gamma^{r_{-}r_{-}} =
\gamma^{ij}D_{i}r_{+}D_{j}r_{+} +\gamma^{ij}D_{i}r_{-}D_{j}r_{-} = 1,
\end{equation}
\begin{equation} \label{g0}
\gamma_{r_{0}r_{0}}=\gamma_{ij}l^{i}l^{j}=
\gamma^{r_{+}r_{+}} \gamma^{r_{-}r_{-}}(r^{2}_{+}-r^{2}_{-}-
\sigma^{-2}_{+}\sigma^{-2}_{-})^{2}.
\end{equation}
Finally, inserting (\ref{b}) into
$(B^{+}_{ij}w^{j}_{-} + B^{-}_{ij}w^{j}_{+}) D^{i}r_{\pm} = 0$
and using (\ref{gr}) and again (\ref{ddw}) and (\ref{kw})
yields linear first order differential equations for
$\gamma^{r_{+}r_{+}}$ or $\gamma^{r_{-}r_{-}}$. Integrating, we find
(\ref{wkerr}) and (\ref{gkerr}) with $r_{+}=r$, $r_{-}=\alpha~ cos\theta$,
$r_{0}=\alpha^{-1} \phi$ and
$\sigma_{\pm}=|m_{\pm}|^{-1/2}$ where $\alpha$ is a constant of integration.
This proves the lemma.\\ \\
{\it Proof of the Theorem.} We prove the Theorem by integrating (\ref{ddkw})
for $\alpha = 1$. Rewriting the l.h. sides in terms of quantities defined above
we find
\begin{equation}
\label{dy}
\nabla_{\mu}[\frac{(1 + {\cal E}_{+})(1 + {\cal E}_{-})}{\lambda}
(\nabla^{\mu}\frac{\sqrt{\mu \pm \nu}}{1- {\cal E}_{\pm}^{2}})
+ \frac{\sqrt{\mu \pm \nu}}{2 \lambda^{2}}(\nabla^{\mu} {\cal E}_{\mp} -
\frac{1 - {\cal E}_{\mp}^{2}}{1 - {\cal E}_{\pm}^{2}}
\nabla^{\mu}{\cal E}_{\pm})] \ge 0,
\end{equation}
and by Lemma 2 equality implies Kerr in the AF case. The vector pair in
brackets, called $Y_{\pm}^{\mu}$, is singular at the nuts and on the sets
${\cal X}_{\pm}$ where $\mu \pm \nu = 0$. The latter are submanifolds of
dimension $\le 2$ and invariant under $\mu_{\tau}$ as can be shown from
(\ref{dde}) like in the static Lor. case \cite{MRS}. We note that at a nut
$\sqrt{\mu \pm \nu} = |\kappa^{+} \pm \kappa^{-}|$, and we assume first that
none of the nuts is (anti-) self dual, viz. $|\kappa^{+} \pm \kappa^{-}|\ne 0$.
Applying the divergence theorem to (\ref{dy}) we get the bounds
\begin{equation}
\label{inty}
0  \le  \int_{\infty} Y_{\mu}^{\pm}dS^{\mu} +
\sum_{i=1,2} \int_{{\cal S}_{i}} Y_{\mu}^{\pm}dS^{\mu} +
\int_{{\cal T}_{\pm}} Y_{\mu}^{\pm}dS^{\mu}
\end{equation}
on surface integrals (with $dS^{\mu}$ directed outwards) over infinity, over
small spheres ${\cal S}_{i}$ around the nuts and over small tubes
${\cal T}_{\pm}$ around ${\cal X}_{\pm}$. Again both bounds are simultaneously
saturated for Kerr only. Performing the limits ${\cal S}_{i} \rightarrow n_{i}$
and ${\cal T}_{\pm} \rightarrow {\cal X}_{\pm}$ as carefully as done in
\cite{MRS} we find that the last pair of integrals in (\ref{inty}) is
non-positive whereas the first two pairs can be evaluated using (\ref{n}) and
Lemma 1. We obtain
\begin{eqnarray}
\label{lpk}
\lefteqn{ 0  \le  -4\pi l_{\infty} \pm  8\pi m_{1}^{\ast}|\kappa_{1}^{+} \pm
\kappa_{1}^{-}| \pm 8\pi m_{2}^{\ast}|\kappa_{2}^{+} \pm \kappa_{2}^{-}| \le}
\nonumber \\
& & \le -4\pi \min(2\pi/\kappa^{+}, 2\pi/\kappa^{-}) +
(4\pi^{2}/\kappa^{+}\kappa^{-})(|\kappa^{+} + \kappa^{-}| -
|\kappa^{+} - \kappa^{-}|) =  \nonumber\\ &  & = 0
\end{eqnarray}
which also follows easily for (and excludes) (anti-) self-dual nuts. This
finishes the proof.\\

Our result can possibly be generalized in various directions. Firstly, it might
be possible to show directly (i.e.  without using results of this paper)
that geodesics emanating from nuts $n_{i}$ with tangent vectors in the
prefered subspaces $T_{n_{i}}^{\pm}$ either join the nuts or reach infinity.
This would yield a stronger version of Lemma 1 (namely  that $l_{\infty}$
equals the period $\tau^{\pm}$ of the corresponding subspaces) without or
under weaker assumptions on the topology of ${\cal M}$.

We also would like to allow "bolts". In fact, we can show as follows that
$({\cal M}, g_{\mu\nu})$ must be the Riem. Schwarzschild metric if ${\cal L}$
is connected. Since the twist scalar satisfies $\nabla_{\mu}(\lambda^{-2}
\nabla^{\mu} \omega) = 0$ which is regular elliptic except at ${\cal L}$,
$\omega$ must
have its maximum and its minimum at ${\cal L}$ or at infinity. But extrema at
the infinity of ${\cal M}$ or ${\cal N}$ can be ruled out by compactifying the
end of ${\cal N}$ (as in the Lor. case \cite{BSK}). Since $\omega$ is constant
on ${\cal L}$ it must vanish identically, i.e. $\xi^{\mu}$ is
hypersurface-orthogonal. The proof can now be completed via any of the Lor.
methods  \cite{IR, MRS, BM}, in particular again by integrating (\ref{dy}).

Of course Lemma 2 suggests that our uniqueness result might be extendable
to the Kerr-NUT case. For this purpose we should assume ALF instead of AF,
 generalize Lemma 1 to include the boundary terms in the signature
\cite{APS, GPR} and note that the dual mass $m^{\ast}$ no longer vanishes.

Furthermore, there presumably result still more general families of
``half-Kerr-NUT'' solutions (and of Lor. counterparts) by
imposing only one of the $''+''$ or $''-''$ parts of (\ref{kw}), (\ref{b0}) or
(\ref{c0}) (or corresponding Lor. equations). Under suitable asymptotic
conditions a uniqueness result for the Riem. solutions might be obtained by
integrating the corresponding part of (\ref{dy}).\\ \\
{\Large\bf Acknowledgement}\\
I am grateful to Lars Andersson, Robert Bartnik, Robert Beig, Piotr
Chru\'sciel, Gary Gibbons and Helmuth Urbantke for helpful
discussions.

\end{document}